\documentclass[preprint,12pt]{elsarticle}
\usepackage{amssymb}
\usepackage{amsmath}
\usepackage{xcolor}
\usepackage{graphicx}
\usepackage[numbers,sort&compress]{natbib}
\usepackage{lineno}
\journal{Combustion and Flame}

\begin{document}

\begin{frontmatter}

\title{Precision Thermometry of Flat Flames Using Spatially Resolved Multi-Color Laser Absorption Spectroscopy of Carbon Dioxide}

\author{Shuoxun Zhang, Shengkai Wang*}

\affiliation{organization={SKLTCS, CAPT, College of Engineering, Peking University},
            addressline={5 Yiheyuan Road, Haidian District}, 
            city={Beijing},
            postcode={100871}, 
            country={China}}
\begin{abstract}
This work developed an accurate and robust absorption-based method for spatially resolved measurements of gas temperatures in flames and reacting flows, with typical single-measurement uncertainties on the order of 1\%. This method exploits narrow-linewidth laser absorption of hot CO$_2$ molecules, which can be generated from combustion or artificially seeded into the flow. A collinear dual-laser setup allowed for periodic scans over tens of CO$_2$ absorption transitions near the $\nu_3$ bandhead every 100 $\mu s$, from which the gas temperature (as well as CO$_2$ concentrations) was determined with high sensitivity and robustness. Spatially resolved measurements were achieved using an electrically driven high-speed beam scanning system consisting of a 2-D galvo scanner and a pair of off-axis parabolic mirrors. An effective spatial resolution of 1 mm was achieved at a planar field measurement speed of 200 Hz and a volumetric field measurement speed of 2 Hz. A physically constrained nonlinear inference framework was also developed for the quantitative analysis of the measurement data. Proof-of-concept experiments were performed on axisymmetric flames stabilized on a Mckenna burner at various equivalence ratios and flow rates, and the results agreed asymptotically with the theoretical value of the adiabatic flame temperature. An additional experiment on a flame of complex geometry demonstrated an excellent level of resolution, precision, and contrast achieved by the current thermometry method. This method promises to provide good utility in future combustion studies due to its high performance metrics and relative ease of use.
\end{abstract}

\begin{highlights}
\item Novelty and Significance\\
This study presented, to the authors' knowledge, the first absorption-based spatially resolved flame thermometry method that achieved a typical single-shot measurement accuracy on the order of 1\%. This method exploited a unique dual-laser setup that allowed rapid scanning over tens of CO$_2$ transitions with drastically different temperature sensitivity, thereby ensuring high measurement sensitivity, accuracy, and robustness. Spatially resolved measurements were achieved using an electrically driven high-speed 2D beam scanning system, with an effective spatial resolution of 1 mm at a planar field measurement speed of 200 Hz and a volumetric field measurement speed of 2 Hz. The performance of this method was thoroughly validated in a series of proof-of-concept experiments. This method is anticipated to be widely adopted in future combustion studies because of its high performance metrics and relative ease of use.

\item Author Contributions\\
Shuocun Zhang: Data curation, Investigation, Methodology, Writing - original draft. \\
Shengkai Wang: Conceptualization, Formal Analysis, Funding acquisition, Investigation, Methodology, Supervision, Writing — original draft, Writing - review \& editing.
\end{highlights}


\end{frontmatter}

\section{Introduction}
Precision measurements of temperature are critically important to modern experimental studies of combustion and reactive flows, as temperature governs local thermal equilibrium, dictates the rates of chemical reactions, and affects the transport of heat, mass, and momentum. Laser-based techniques are particularly useful in this regard due to their capabilities to measure temperature with high spatial and temporal resolution in a quantitative and non-intrusive manner. Some of the most widely used techniques include linear methods such as laser absorption spectroscopy (LAS) \cite{hanson1978temperature, zhou2005development, rieker2009calibration, cassady2021time, mathews2023experimental}, laser Rayleigh/Raman scattering spectroscopy \cite{dibble1981laser, fourguette1986two, bohlin2014diagnostic, kristensson2015advancements} and laser-induced fluorescence (LIF) \cite{seitzman1985instantaneous, mcmillin1993temporally, palmer1996temperature, lee2005quantitative, devillers2008development, wang2019quantitative}, as well as nonlinear methods such as coherent anti-Stokes Raman scattering (CARS) \cite{eckbreth1980cars, lucht1987unburned, hancock1997nitrogen, kearney2009dual, cantu2018temperature, athmanathan2021femtosecond}, degenerate four-wave mixing (DFWM) and polarization spectroscopy \cite{nyholm1994single, klamminger1995rotational, sun2011flame, sahlberg2019mid, song2024temperature}. The merits and drawbacks of each method have been discussed in several comprehensive reviews, such as \cite{laurendeau1988temperature, hanson1988combustion, wolfrum1998lasers, stricker2002measurement, kohse2005combustion, roy2010recent, hanson2011applications, alden2011visualization, kiefer2011laser, goldenstein2017infrared, alden2023spatially}. A prevailing challenge to current thermometry methods is the trade-off between precision, accuracy, and robustness in relation to spatial/temporal resolution and measurement complexity. For example, absorption techniques are quantitative and robust but often lack spatial resolution along the laser propagation direction; fluorescence methods can provide instantaneous 2D measurements but may suffer from J-dependent quenching \cite{copeland1985rotational, kienle1996detailed, yan2024star} and radiation trapping effects \cite{lucht1982temperature, sadanandan2012experimental}; non-resonant Rayleigh and Raman scattering methods do not require specific excitation wavelengths but have low signal yields; and nonlinear optical techniques are very sensitive but experimentally complicated and often require laborious data evaluation. New developments and improvements are needed, particularly for accurate and simple thermometry methods with good spatial and temporal resolution.

Of particular interest to the current study are thermometry methods based on laser absorption spectroscopy, which are especially useful in situations where high levels of measurement accuracy are demanded, such as in studies of reaction kinetics. It was estimated that a 1\% change in gas temperature would modify the overall reactivity of most combustible mixtures by more than 10\%, based on ignition delay time measurements of 30 representative fuels \cite{davidson2004interpreting, davidson2017ignition}.  Similar levels of accuracy are required in studies of transport phenomena, where small temperature differences on the order of tens of Kelvin need to be resolved for good gradient measurements. To date, reducing temperature measurement uncertainty below 1\% remains a challenging task; even for the simple configuration of premixed laminar one-dimensional flame above a flat burner, the level of accuracy is still much less than ideal. For example, the reported single-shot measurement uncertainty of flame temperature generally ranges from 2 to 5\% for LAS \cite{cassady2021time, spearrin2014simultaneous, ma2018situ}, from 2 to 10\% for LIF \cite{lee2005quantitative, wang2019quantitative, hanson1988combustion, seitzman1994application, bessler2004quantitative}, and is typically higher for CARS and DFWM (although their uncertainties can be reduced by temporal averaging). A close examination of the classic LAS and LIF thermometry methods (mostly two-color/two-line methods) revealed that most of the uncertainty results from the intensity ratio measurements and that there appeared to be an inherent contradiction between the measurement sensitivity and robustness -- high sensitivity can amplify the noise and systematic error in the intensity of the weaker transition and spoil the measurement. To achieve good accuracy, extreme care was required to characterize the laser intensity baseline, suppress flame emission/chemiluminescence, and eliminate interfering absorption/fluorescence. This challenge is often compounded by the lack of an in situ calibration standard under flame conditions, where conventional temperature references such as resistance thermometers and thermocouples fail at such high temperatures.

The key to improving temperature measurement accuracy, as recognized in several previous studies (e.g., \cite{sanders2001diode, bessler2004quantitative,  lin2010selection, ma2017non, malarich2021resolving}) as well as in the present work, lies in the ability to access multiple transitions of different lower-state energies and to leverage the measurement uncertainties between them. An exemplary series of research along this path is the recent advancement of frequency comb spectroscopy for combustion diagnostics, which has enabled massive parallel detection of hundreds of molecular transitions \cite{schroeder2017dual, makowiecki2021mid, schroeder2021temperature, zhu2023mid}. However, the power density per comb tooth was too low for high-quality measurements in a single-shot manner, and long-time averaging was usually required to achieve a decent signal-to-noise ratio.  In this regard, the current work explored an alternative method using collinear beams of narrow-linewidth lasers to access tens of closely spaced molecular transitions in a time-multiplexed fashion. This method exploits the high spatial and spectral power density of narrow-linewidth lasers and can be advantageous for measurements in harsh environments where strong emissions and rapid temporal variations are present. 

Another critical challenge for LAS thermometry lies in its limited spatial resolution, but this problem can be solved, at least in part, by adding measurements along different lines of sight. This very concept has evolved into a rich family of spectroscopic measurement techniques known as laser absorption tomography, with notable examples including the studies of \cite{ma2009tomographic, ma201350, cai2017tomographic, grauer2023volumetric}. In spite of these successful applications, however, conventional tomography required the use of a large number of lasers, which was laborious and not cost-effective. Recently, 2D cinematographic imaging combined with tunable Mid-IR laser illumination has provided an alternative path to spatially resolved measurements of gas temperature \cite{tancin20192d, wei2021volumetric, wei2023four}. This method avoids the use of many lasers and substantially boosted the measurement efficiency, but at the cost of reduced power density and limited wavelength tuning range, which degrades the measurement accuracy. In light of these issues, the current study has also explored a time-multiplexed version of 2D absorption imaging, with collimated laser beams probing a single line of sight at a time.

In summary, the primary objective of the current study was to develop an accurate and robust absorption-based method for spatially resolved measurements of gas temperatures in flames and reacting flows. In conjunction with this method, a constrained nonlinear tomographic inference framework was also introduced to analyze the measurement data quantitatively. The remainder of this paper is organized as follows: Section 2 elaborates on the details of the current experimental methods; Section 3 introduces the data analysis methods; Section 4 presents a series of proof-of-concept experiments and evaluates the performance metrics of the current thermometry method. It is worth noting that the current study has also resolved the issue of validation/reference standards at flame conditions via asymptotic analysis. Section 5 concludes this work with a brief summary and outlook.

\section{Experimental Methods}
\label{Experimental Methods}

\begin{figure}[ht!]
\centering
\includegraphics[width=\linewidth]{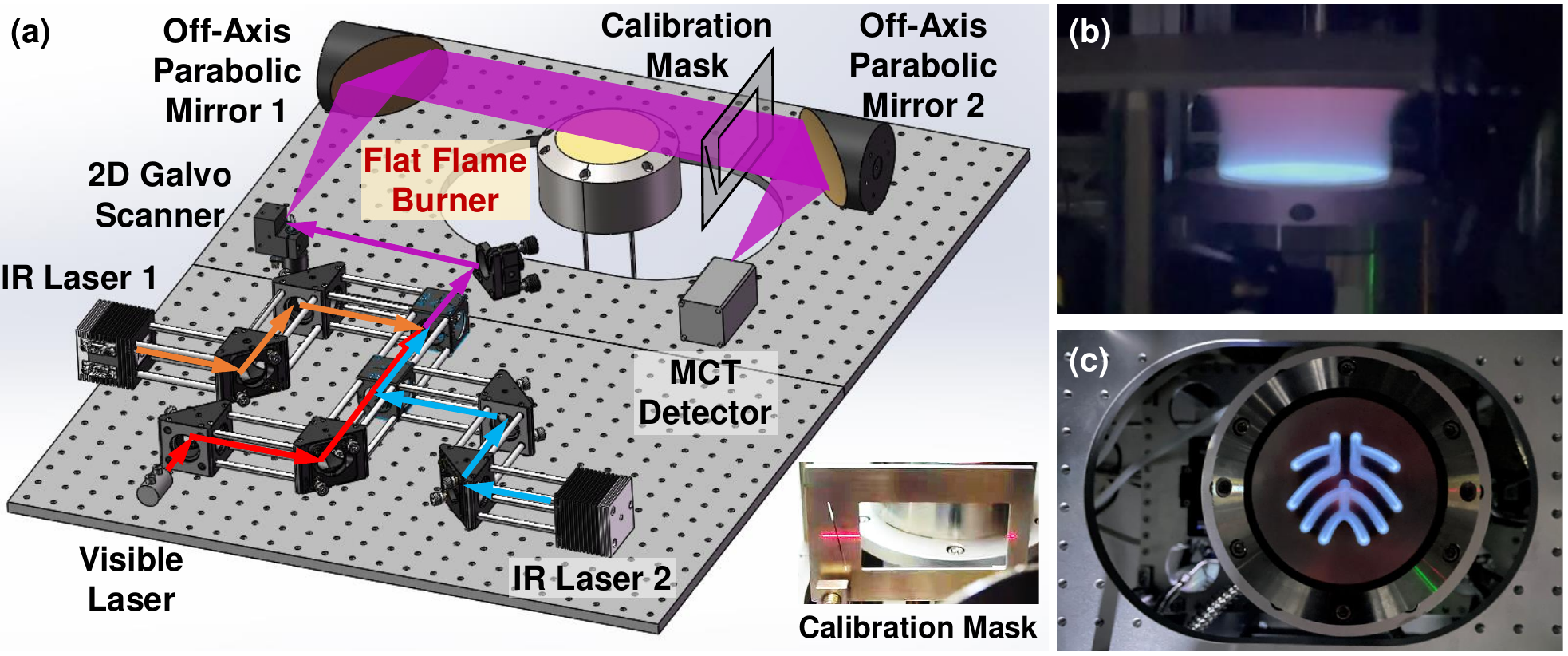}
\caption{A schematic of the current experimental setup. Panel (a): laser diagnostics system. Panel (b): axisymmetric flame with a stagnation plate. Panel (c): PKU flame.}
\label{Fig1}
\end{figure}

\subsection{Laser Absorption Diagnostics}
The current thermometry method used narrow-linewidth lasers to probe 27 absorption transitions of hot CO$_2$ molecules near the $\nu_3$ bandhead. CO$_2$ was selected as the tracer species due to its relative inertness; it is naturally present in hydrocarbon flames as one of the major combustion products, but can also be artificially seeded into the flow. As shown in Fig. \ref{Fig1}(a), these absorption features were probed by two Nanoplus distributed feedback interband cascade lasers (DFB-ICL) operated near 4175 nm and 4179 nm, respectively. The output beams of these two lasers were aligned collinearly using 1-inch cage-mount optics, ensuring that they always measured the same volume of gas. The lasers were carefully collimated using anti-reflection-coated 1/2-inch ZnSe lenses, and their beam waists were positioned near the flame center. The $1\sigma$ diameters of both beams were measured to be 1.0 $\pm$ 0.1 mm at the waists using an IR laser beam profiler, and their Rayleigh ranges were on the order of 0.6 m. As a visual aid for the optical alignment and a visible tracker of the measurement beam path, a 650-nm collimated diode laser was also added to the collinear beam setup. The $1\sigma$ beam diameter of the visible laser was approximately 0.5 mm, and the angular misalignment between these three beams was less than 0.3 mrad, as estimated by photodetector measurements 2 m downstream of the beam waists.

The use of a dual-laser setup allowed for versatile access to a large number of absorption transitions with various temperature sensitivities. The two lasers were time-multiplexed by periodic current modulations at 10 kHz, and a constant phase offset between the modulations ensured that only one laser was above its threshold at any time. This current modulation was precisely controlled by two ppqSense QubeDL02-T laser controllers that were synchronized to a RIGOL DG2052 16-bit, 2-channel function generator. The relative wavenumber change of each laser during a current modulation cycle was characterized by a solid 3-inch germanium etalon with an FSR of 0.0164 $\rm{cm^{-1}}$. Within each modulation cycle of 100 $\rm{\mu s}$, the 4175-$\rm{nm}$ laser was rapidly scanned over the R(102) - R(142) transitions between 2395.85 - 2397.50 $\rm{cm^{-1}}$ during the first 58 $\rm{\mu s}$, whereas during the last 42 $\rm{\mu s}$, the 4179-nm laser was scanned over the R(84) - R(88) and R(156) - R(160) transitions between 2392.45 - 2393.80 $\rm{cm^{-1}}$. From the relative intensities of these absorption transitions, the gas temperatures and CO$_2$ concentrations were determined with high sensitivity and robustness. 

In addition to the collinear dual-laser setup, the current study also employed a 2D high-speed parallel beam scanning system for spatially resolved measurements (see Fig. \ref{Fig1}). This beam scanning system consisted of a Thorlabs GVS102 2D galvo scanner and a pair of off-axis parabolic mirrors (focal length = 9 inches, projected diameter = 3 inches) placed upstream and downstream of the measurement region. The galvo scanner had two orthogonal mirrors that were electrically modulated by another RIGOL DG2052 function generator at 100 Hz (horizontal) and 1 Hz (vertical), respectively, and the resulting oscillations of the mirrors swept the reflected beam angle by $\pm$ 12.5 degrees horizontally and up to 7.5 degrees vertically. The output mirror of the galvo scanner was located right at the focus of the upstream off-axis parabolic mirror; therefore, the reflected beams off the parabolic mirror were always parallel to its axis. The second off-axis parabolic mirror focused the parallel beams onto a VIGO PVI-4TE-5 MCT detector. A narrow bandpass filter (Spectrogon NB-4180-074, with a center wavelength of 4180 nm and FMHW of 74 nm) was placed in front of the detector to block the IR emission of the flames while transmitting most of the laser intensity ($\geq$ 70\%). The average transmitted laser power was approximately 3 mW, which was above the saturation intensity of the detector; therefore, a neutral density filter was also added. 

In essence, the current beam scanning scheme was a time-multiplexed version of 2-D absorption imaging, with collimated laser beams probing a single line of sight at a time. The transmitted laser intensity and the horizontal modulation waveform of the galvo scanner were digitally recorded using a NI PXI-5122 high-speed data acquisition module at a rate of 100 MS/s. For each beam scanning cycle, the horizontal position of the laser beam was determined from the recorded modulation waveform, while its vertical position was determined using a specially designed mask (see the bottom corner of Fig. \ref{Fig1}(a)) that acted as a spatial encoder. Since each location in the measurement region was traversed twice per scan cycle, a planar measurement speed of 200 Hz and a volumetric measurement speed of 2 Hz were routinely achieved. Further improvement in the measurement speed by another order of magnitude was attainable, as indicated by tests at galvo frequencies up to 1 kHz, although additional calibrations were required to correct the nonlinear distortions of the galvo scanner under high centrifugal forces.

\subsection{Flat-Flame Burner Configuration}
To evaluate the accuracy and precision of the current thermometry method, a series of experiments were conducted on premixed CH$_4$-air flames that were stabilized on a water-cooled McKenna burner, as shown in Fig. \ref{Fig1}(b). These flames were axisymmetric and approximately 60 mm in diameter, and they were shielded from the ambient air by an annular nitrogen flow of approximately 10 mm thick. This nitrogen shielding also reduced the shear between the flame and the surrounding air. A stagnation plate made of fused quartz was placed 30 mm above the flames to enhance stability, especially for experiments at high flow rates where the flame tended to blow off. Research-grade high-purity $\rm CH_4$ (99.99\%-grade) and synthetic air (prepared from 99.999\%-grade $\rm N_2$ and 99.999\%-grade $\rm O_2$) were supplied to an inline static mixer before entering the burner. A stream of high-purity $\rm N_2$ was supplied separately to the shielding flow. The flow rates of $\rm CH_4$, air, and $\rm N_2$ were precisely controlled by three Alicat MC series mass flow controllers, with typical uncertainties of 0.1\%, 0.2\%, and 0.5\%, respectively. 

To demonstrate the resolution and contrast of the current thermometry method, an additional experiment was also conducted on a flame of complex geometry. This flame, referred to as the PKU flame in the current study, was generated by masking the McKenna burner with a 5-mm-high stainless-steel logo of Peking University, as shown in Fig. \ref{Fig1}(c). It was a premixed $\rm C_3H_8$-air flame without nitrogen shielding, and the mass flow rates of high-purity $\rm C_3H_8$ (99.95\%-grade) and synthetic air were fixed at 0.40 SLPM (standard liter per minute) and 9.61 SLPM, respectively, corresponding to an equivalence ratio of 1.00. Since the PKU flame was not axisymmetric, measurements at multiple viewing angles were required. This was achieved by placing the entire burner on an electrically controlled high-precision rotation stage with an angular uncertainty of less than 5 $\rm{\mu rad}$. All tubes for the gases and the cooling water were made flexible so that they would not impede the rotation of the burner.

\section{Data Analysis}
\subsection{The Overall Framework}
\begin{figure}[ht!]
\centering
\includegraphics[width=\linewidth]{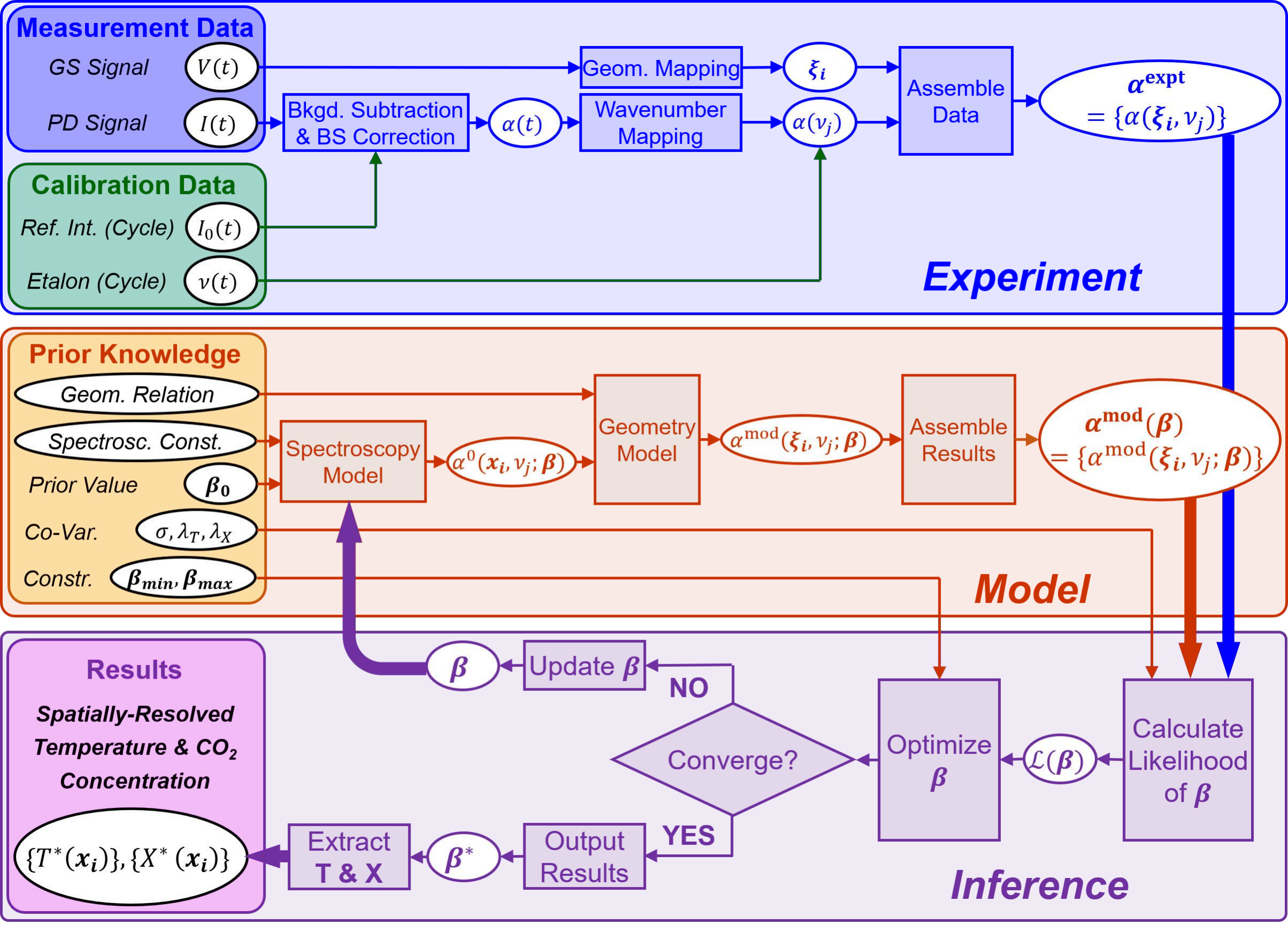}
\caption{A schematic diagram of the current framework for analyzing the absorption data}
\label{Fig2}
\end{figure}

As illustrated in Fig. \ref{Fig2}, a physically constrained nonlinear inference framework was developed in the current study for the quantitative analysis of the measurement data. This framework consisted of three interconnected modules: (1) an experiment data reduction module to extract the absorption signal, (2) a forward modeling module to simulate the absorption spectra at various lines of sight based on prior knowledge and trial distributions of the gas properties, and (3) an inference module to iteratively determine the most probable values of gas properties across the measurement region via constrained optimization. The details of these modules are explained in the following subsections.

\subsection{Extracting the Absorption Signal}
\begin{figure}[ht!]
\centering
\includegraphics[width=0.5\linewidth]{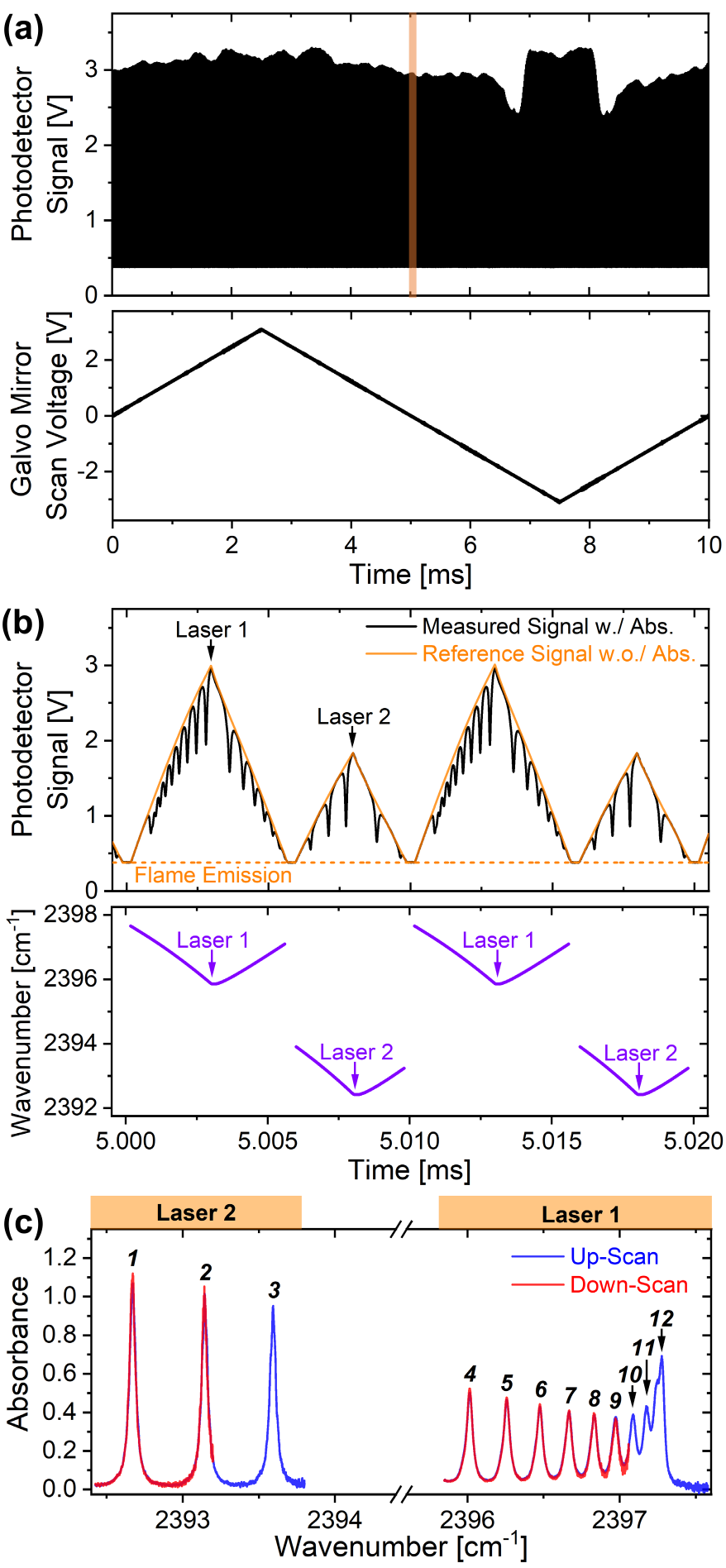}
\caption{Example data obtained in the current study. Panel (a): raw signals of the photodetector (top) and the galvo scanner modulation (bottom). Panel (b): a zoomed-in view of the region highlighted in (a) illustrating the intensity (top) and wavenumber (bottom) modulations of the lasers. Panel (c): the absorbance signal extracted from the raw data within one modulation cycle of 100 $\rm{\mu s}$.}
\label{Fig3}
\end{figure}

As shown in the top part of Fig. \ref{Fig2}, the general procedure for extracting the absorption signal in a flame experiment involves several steps and requires four types of data inputs: the photodetector (PD) signal $I(t)$ (with absorption), the galvo scanner (GS) modulation signal $V(t)$, the reference intensity of the lasers $I_0(t)$ (without absorption) within a current modulation cycle of 100 $\rm{\mu s}$, and the etalon signal of laser wavenumber modulation $\nu(t)$ within a cycle. This procedure is further illustrated in Fig. \ref{Fig3} by a representative set of data obtained from the axisymmetric flame experiment.

First, background subtraction and beam-steering correction were applied to $I(t)$ cycle-by-cycle via comparison with the reference signal $I_0(t)$, as shown in Fig. \ref{Fig3}(b). The basic concept of this step was to shift and rescale $I(t)$ in each individual cycle to eliminate the interference of flame emission (which added a nearly constant baseline) and intensity variations due to beam steering (which modified the received laser intensity by nearly a constant ratio). This was achieved through a simple linear regression of $I(t)$ with respect to $I_0(t)$ at wavelengths where the lasers were below threshold and/or where the $\rm CO_2$ absorption was negligible, and any residual error in the form of a small absorbance offset was corrected iteratively based on a spectral fit. After the correction of $I(t)$, the absorbance signal $\alpha(t)$ was then analyzed using the Beer-Lambert relation, as described in Eqn. \ref{Eqn1}.
\begin{equation} \label{Eqn1}
\begin{aligned}
\alpha(t) & = \ln \big[I(t)/I_0(t)\big]
\end{aligned}
\end{equation}

Next, the time sequence of the absorbance signal was mapped to the wavenumber space based on the $\nu(t)$ data shown in Fig. \ref{Fig3}(b). This mapping translated $\alpha(t)$ in each laser modulation cycle to an absorbance spectrum $\alpha(\nu_i)$, as displayed in Fig. \ref{Fig3}(c). Note that many transitions were scanned twice during each cycle (i.e. in the up-scan and the down-scan); this served as a quick check for errors and as a good reference to evaluate the consistency and quality of the reduced data. 

In addition, the galvo scanner signal $V(t)$ was also translated into beam position parameters $\boldsymbol{\xi_i}$ based on the geometric mapping described in Section 2.1. The exact format of $\boldsymbol{\xi_i}$ depended on the nature of the experiment. For example, in a planar measurement of an axisymmetric flame, the beam position was fully characterized by a single parameter, such as the distance from the flame center. In contrast, for an asymmetric flame and/or a volumetric measurement, additional parameters such as the azimuth angle and/or the height above burner were required.

Lastly, the absorbance data extracted at different wavenumbers and beam positions were combined into a single column vector represented as $\boldsymbol{\alpha^{expt}} = \{ \alpha(\boldsymbol{\xi_i},\nu_j) \}$.

\subsection{Modeling the Absorption Spectra}
The absorption spectra of $\rm CO_2$ were modeled using the latest version of the HITEMP database. First, the local absorbance spectrum $\alpha^0(\boldsymbol{x_i},\nu_j)$ at location $\boldsymbol{x_i}$ was calculated as a function of the local gas properties (including temperature $T_i$, pressure $P_i$, $\rm CO_2$ mole fraction $X_i$, etc.; denoted collectively as $\boldsymbol{\beta_i}$) based on Eqn. \ref{Eqn2}. 
\begin{equation} \label{Eqn2}
\begin{aligned}
\alpha^0(\boldsymbol{x_i},\nu_j;\boldsymbol{\beta_i}) & = \alpha^0(\boldsymbol{x_i},\nu_j;T_i,X_i,P_i,\{\Delta\nu_{C,ki}\},\{\delta\nu_{0,ki}\}) \\
& = X_i P_i \sum_{k = 1}^{K} S_k(T_i) \phi_V(\nu_j - \nu_{0,k} - \delta\nu_{0,ki}, \Delta\nu_{C,ki}, \Delta\nu_{D,ki}) \\
\end{aligned}
\end{equation}

This equation accounted for the contributions from a total number of $K$ adjacent absorption transitions, with $S_k$, $\nu_{0,k}$, $\Delta\nu_{D,ki}$, $\Delta\nu_{C,ki}$, and $\delta\nu_{0,ki}$ representing the linestrength, center wavenumber, Doppler linewidth (FWHM), collisional linewidth (FWHM), and collision shift of the $k$-th transition at $\boldsymbol{x_i}$, respectively. In the current study, these transitions were modeled with the Voigt lineshape function $\phi_V$, which was evaluated numerically using the Humlicek \cite{humlivcek1982optimized} algorithm. Despite the somewhat complex form of $\phi_V$, its integral over wavenumber was always equal to unity, i.e.,
\begin{equation} \label{Eqn3}
\begin{aligned}
\int_{0}^{+\infty} \phi_V(\nu - \nu_{0,k} - \delta\nu_{0,ki}, \Delta\nu_{C,ki}, \Delta\nu_{D,ki}) d\nu  = 1\\
\end{aligned}
\end{equation}

Therefore, for strong transitions that are sufficiently separated in wavenumber space, their integrated absorbances $A_k^0(\boldsymbol{x_i})$ depend primarily on $T_i$, $P_i$, and $X_i$ only, i.e.,
\begin{equation} \label{Eqn4}
\begin{aligned}
A_k^0(\boldsymbol{x_i};T_i,X_i,P_i) & = X_i P_i S_k(T_i) \\
\end{aligned}
\end{equation}
And for atmospheric flames, the pressure dependence disappeared as $P_i$ = $P_0$ = 1 $atm$. In the current study, this spectral integration approach was employed to simplify the analysis of absorption features 1-9 in Fig. \ref{Fig3}(c), with slight modifications to account for the overlapping weak transitions and the absorption wings of adjacent transitions. This was achieved by conducting a lineshape-resolved calculation (normalized by $X_i$ and $P_0$) using an estimated value of $\Delta\nu_{C,ki}$, and integrating the absorbance over a designated wavenumber interval around each absorption feature. The resulting effective linestrength was characterized as a function of the local gas temperature, i.e., $\Tilde{S_k^0}(T_i)$. The integrated absorbance $\Tilde{A_k^0}(\boldsymbol{x_i})$, defined over the same wavenumber interval as $\Tilde{S_k^0}$, depended on $T_i$ and $X_i$ only, as given by
\begin{equation} \label{Eqn5}
\begin{aligned}
\Tilde{A_k^0}(\boldsymbol{x_i};T_i,X_i) & = X_i P_0 \Tilde{S_k^0}(T_i) \\
\end{aligned}
\end{equation}

The local absorbance $\alpha^0$ was then spatially integrated along the line of sight specified by $\boldsymbol{\xi_i}$, based on the geometry model, as shown in Eqn. \ref{Eqn6}. 
\begin{equation} \label{Eqn6}
\begin{aligned}
\alpha^{mod}(\boldsymbol{\xi_i},\nu_j;\boldsymbol{\beta}) & = \int_{-\infty}^{+\infty} \alpha^0\Big(\boldsymbol{x(\xi_i},l), \nu_j; \boldsymbol{\beta\big(x(\xi_i},l)\big)\Big) dl \\
&\approx \sum_{s = 1}^{\# pts.} w_{is} \cdot \alpha^0(\boldsymbol{x_s},\nu_j;\boldsymbol{\beta_s})
\end{aligned}
\end{equation}

Evaluation of this integral requires the gas properties $\boldsymbol{\beta(x)}$ to be defined continuously along the beam path $\boldsymbol{x(\xi},l)$. In the current study, this was achieved through spatial interpolation with local shape functions -- a method commonly used in finite element analysis -- that approximated the integral by a weighted sum of the absorbances at selected node points, $\alpha^0(\boldsymbol{x_s},\nu_j; \boldsymbol{\beta_s})$. A similar interpolation approach has been documented by Grauer et al. in their previous study \cite{grauer2017measurement}, and readers interested in spatial interpolation methods are also referred to \cite{schweiger1993application} and \cite{grauer2023volumetric}. Here, extensive details about the formulation of shape functions have been omitted for brevity, but a few key points in the current study are highlighted as follows. (1) The node points $\boldsymbol{x_s}$ were selected based on the effective spatial resolution of the laser diagnostics, which was 1 $mm$, as defined by the diameters of the laser beams. (2) The shape functions were kept simple for robustness. For planar measurements of the PKU flame and volumetric measurements of the axisymmetric flames, 2-D bilinear shape functions with 4 nodes were used, whereas for planar measurements of axisymmetric flames, 1-D 2-node linear shape functions were sufficient. (3) The weight factors $w_{is}$ were geometric constants independent of the laser wavenumber $\nu_j$ and/or the local gas properties $\boldsymbol{\beta_s}$. Therefore, the same weight factors applied to the spectrally integrated absorbance $\Tilde{A_k^0}$ as well.

Lastly, the simulated absorbances were assembled into two column vectors, $\boldsymbol\alpha^{0}(\boldsymbol{\beta}) = \{ \alpha^{0}(\boldsymbol{x_i},\nu_j;\boldsymbol{\beta_i}) \}$ and $\boldsymbol\alpha^{mod}(\boldsymbol{\beta}) = \{ \alpha^{mod} (\boldsymbol{\xi_i},\nu_j;\boldsymbol{\beta}) \}$, with $\boldsymbol{\beta}$ denoting $\{{\boldsymbol{\beta_s}}\}$ collectively. The spectrally integrated absorbances were assembled into column vectors of $\boldsymbol{A^{0}(T,X)} = \{A^{0}(\boldsymbol{x_i};T_i,X_i)\}$ and $\boldsymbol{A^{mod}(T,X)}$ in a similar way, whereas all the weight factors $\{w_{is}\}$ were compiled into a constant matrix $\boldsymbol{W}$. This led to

\begin{equation} \label{Eqn7}
\begin{aligned}
\boldsymbol\alpha^{mod}(\boldsymbol{\beta}) = \boldsymbol{W} \boldsymbol\alpha^{0}(\boldsymbol{\beta})
\end{aligned}
\end{equation}

\begin{equation} \label{Eqn8}
\begin{aligned}
\boldsymbol{A^{mod}(T,X)} = \boldsymbol{W A^0(T,X)}
\end{aligned}
\end{equation}

\subsection{Constrained Tomographic Reconstruction}
The gas temperatures and $\rm CO_2$ concentrations across the measurement region, along with other gas properties in $\boldsymbol{\beta}$, can be determined by solving the constrained optimization problem shown in Eqn. \ref{Eqn9}.
\begin{equation} \label{Eqn9}
\begin{aligned}
\boldsymbol{\beta^*} = \underset{[\boldsymbol{\beta_{min},\beta_{max}}]}{\operatorname{argmin}} \Big\| \boldsymbol{ \Sigma_{\alpha}^{-1/2} [\alpha^{expt} - \alpha^{mod}(\beta)] }  \Big\|_{2}^{2} + \Big\| \boldsymbol{ \Sigma_{\beta}^{-1/2} (\beta - \beta_0)} \Big\|_{2}^{2}
\end{aligned}
\end{equation}

This equation was derived from a Bayesian perspective. The objective function was the negative logarithm of the posterior probability distribution function (PDF) of $\boldsymbol{\beta}$, where $\boldsymbol{\Sigma_{\alpha}}$ and $\boldsymbol{\Sigma_{\beta}}$ represented the covariance matrices of the absorbance measurements $\boldsymbol{\alpha^{expt}}$ and the prior PDF of $\boldsymbol{\beta}$, respectively. $\boldsymbol{\beta_{min}}$ and $\boldsymbol{\beta_{max}}$ denoted the lower and upper limit constraints, while $\boldsymbol{\beta_0}$ denoted the prior mean.

An alternative approach was to infer the temperatures and $\rm CO_2$ concentrations based on the spectrally integrated absorbances, $\boldsymbol{A^{expt}}$, as shown in Eqn. \ref{Eqn10}. This approach represented a simplified version of Eqn. \ref{Eqn9}, where the covariances were assumed to be in the form of an identity matrix or a discretized Laplacian matrix multiplied by some constant. Since scaling the objective by a constant does not change the optimal point, only two constants, $\lambda_T$ and $\lambda_X$, were used to characterize the covariances. This method, known as Tikhonov regularization in several previous studies, was reinterpreted here within a Bayesian framework. The values of $\lambda_T$ and $\lambda_X$ were generally kept small to avoid blurring the reconstructed distribution at any visible level, while being large enough to ensure numerical stability and suppress random fluctuations.

\begin{equation} \label{Eqn10}
\begin{aligned}
(\boldsymbol{T^*,X^*}) = 
\underset{\begin{subarray}{c}
  [\boldsymbol{T_{min},T_{max}}] \\
  [\boldsymbol{X_{min},X_{max}}]
  \end{subarray}}
{\operatorname{argmin}} \big\| \boldsymbol{ A^{expt} - A^{mod}(T,X)} \big\|_{2}^{2} + \lambda_{T}^2 \big\| \boldsymbol{LT} \big\|_{2}^{2} + \lambda_{X}^2 \big\| \boldsymbol{LX} \big\|_{2}^{2}
\end{aligned}
\end{equation}

The current study combined the two approaches by iteratively solving Eqns. (9) and (10) in succession, achieving good computational efficiency, accuracy, and robustness at the same time. The iteration began by numerically solving Eqn. (10) using the interior point algorithm, based on an initial estimate of $\{\Delta\nu_{C,ki}\}$ and $\{\delta\nu_{0,ki}\}$ (which was used to evaluate the integrated linestrengths over finite intervals). Their values were then updated by solving Eqn. (9), while $\boldsymbol{T}$ and $\boldsymbol{X}$ were kept unchanged. After that, Eqn. (10) was solved again, and this process was repeated until convergence was reached, usually within 2 or 3 iterations. To further enhance the stability of the current numerical algorithm, a preconditioning step was applied via a linear transform of the variables, i.e., scaling the physical variables $\boldsymbol{\beta}$, $\boldsymbol{T}$, and $\boldsymbol{X}$ by appropriate constants so that the derivatives of the objective function with respect to the scaled variables were of similar numerical magnitude. The values of the original physical variables were easily recovered via inverse scaling.

\section{Results and Discussion}
\subsection{Planar Thermometry of Axisymmetric Flat Flames}
\begin{figure}[ht!]
\centering
\includegraphics[width=0.5\linewidth]{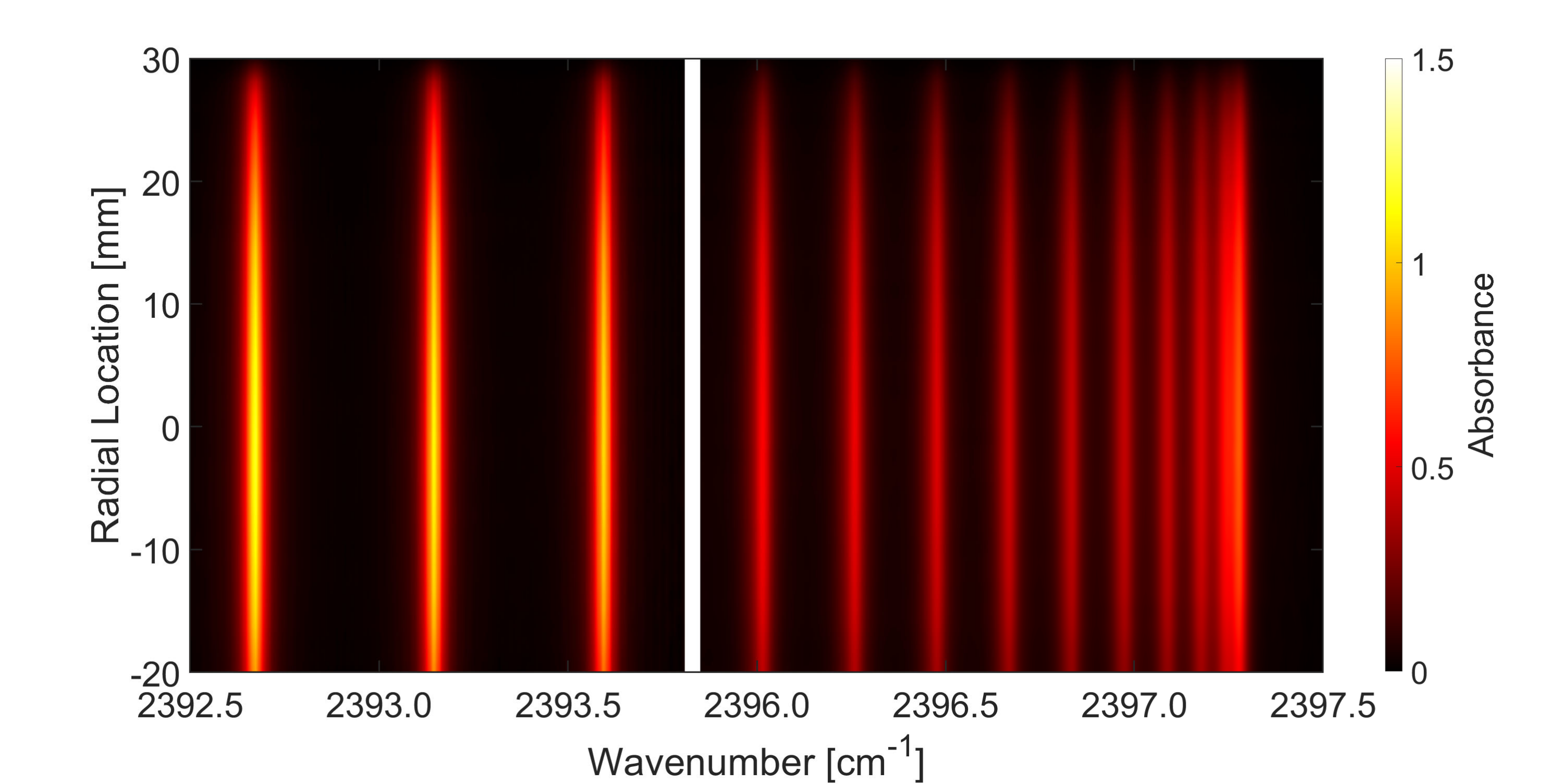}
\caption{Absorbance data obtained from a single-shot planar measurement of an axisymmetric flame at mass flow rates of $\dot{m}_{CH_4}$ = 2.50 SLPM, $\dot{m}_{air}$ = 24.15 SLPM, and $\dot{m}_{N_2}$ = 28.17 SLPM, with an equivalence ratio of $\phi$ = 1.00. The measurement was taken at a height of 5 mm above the burner surface.}
\label{Fig4}
\end{figure}

Fig. \ref{Fig4} shows a representative set of absorbance data, $\boldsymbol{\alpha^{expt}} = \{ \alpha(\boldsymbol{\xi_i}, \nu_j)\}$, obtained from a single-shot planar measurement of an axisymmetric flame conducted within 0.01 s. A typical noise level in the absorbance data was estimated to be $\sigma = 4 \times 10^{-3}$, based on a non-absorption reference experiment. The covariance matrix of $\boldsymbol{\alpha^{expt}}$ was calculated as $\boldsymbol{\Sigma_{\alpha}} = \sigma^2 \boldsymbol{I}$ (where $\boldsymbol{I}$ is the identity matrix), assuming that the noise in each element of $\boldsymbol{\alpha^{expt}}$ can be modeled as independent and identically distributed (IID) random variables.

\begin{figure}[ht!]
\centering
\includegraphics[width=0.5\linewidth]{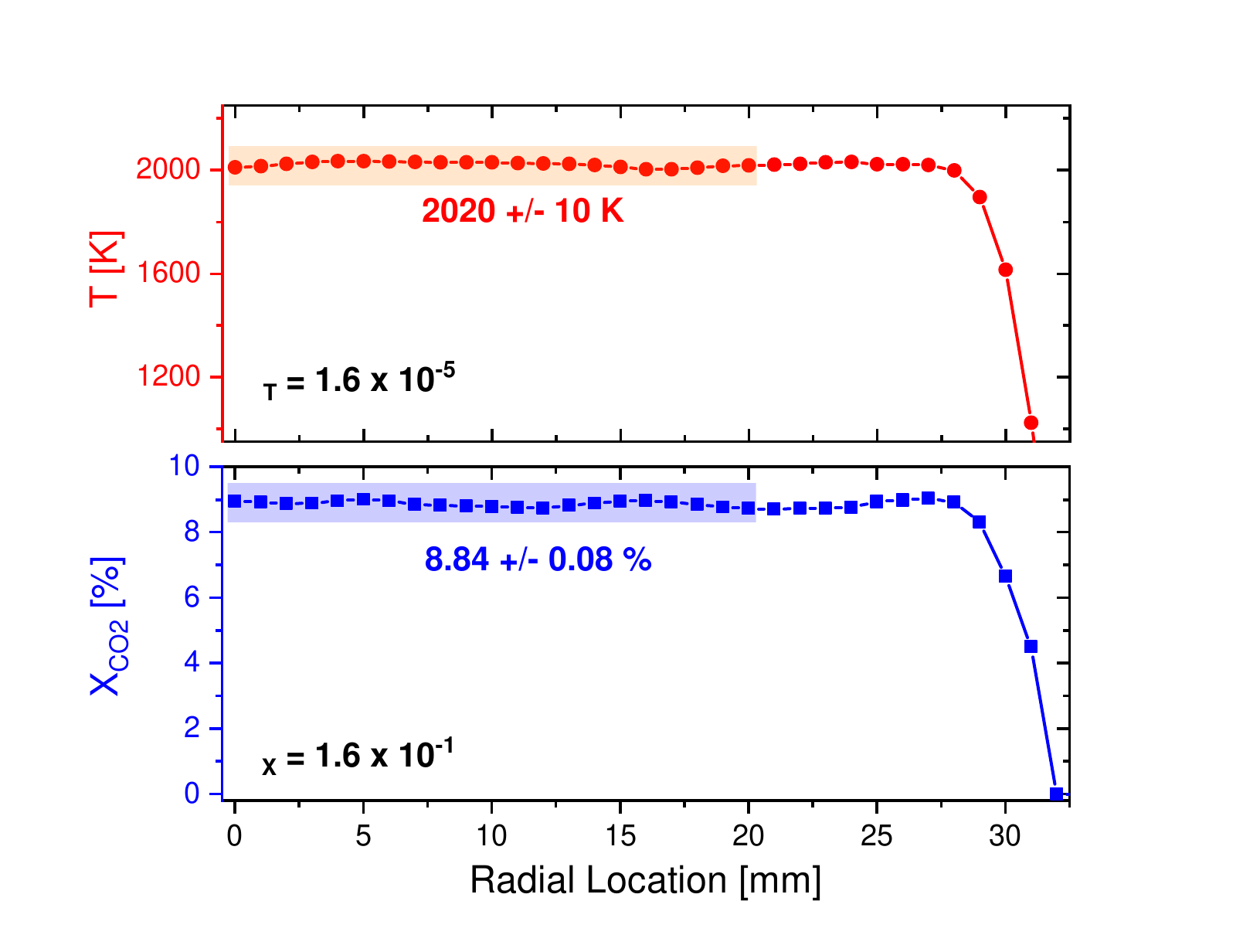}
\caption{The radial distributions of gas temperature and $\rm CO_2$ mole fraction inferred from the absorbance data shown in Fig. \ref{Fig4}. The measurement was taken at a height of 5 mm above the burner surface.}
\label{Fig5}
\end{figure}

The spatial distributions of temperature ($\boldsymbol{T}$) and $\rm CO_2$ mole fraction ($\boldsymbol{X}$), determined using the iterative approach described in Section 3.4, are shown in Fig. \ref{Fig5}. In the current study, only the first 9 of the 12 transition features (up to 2397.1 $\rm{cm^{-1}}$) were analyzed to avoid complications from overlapping transitions and potential line mixing effects. The values of $\lambda_T$ and $\lambda_X$ were set to be $1.6 \times 10^{-5}$ and $1.6 \times 10^{-1}$, respectively, and uniform prior PDFs were assumed for $\{\Delta\nu_{C,ki}\}$ and $\{\delta\nu_{0,ki}\}$. Note that in this study, the inferred $\boldsymbol{T}$ and $\boldsymbol{X}$ were found to be insensitive to the values of $\lambda_T$ and $\lambda_X$, especially in the core region of the flame (the results changed by less than $\pm 0.5\%$ when $\lambda_T$ and $\lambda_X$ varied by an order of magnitude). Further discussion
on the effect of $\lambda_T$ and $\lambda_X$ can be found in the Supplementary Material.

The $1\sigma$ uncertainty in the inferred local gas temperature was estimated from the contributions of individual sources. It was found that random noise in the absorbance measurement contributed very little (less than $0.1\%$), as it was averaged over more than 7,500 different wavenumbers ($\nu_j$) within each laser modulation cycle of 100 $\rm{\mu s}$. Systematic errors in the Tikhonov regularization, linestrength calculation, and background correction, on the other hand, were the dominant sources of uncertainty. Except for a few points near the edge of the flame, their contributions to the temperature uncertainty were relatively small, estimated to be 0.5\%, 0.6\%, and 0.6\%, respectively. The overall $1\sigma$ uncertainty, calculated on a root-sum-squared basis, was $\pm 1.0\%$.

\begin{figure}[ht!]
\centering
\includegraphics[width=0.5\linewidth]{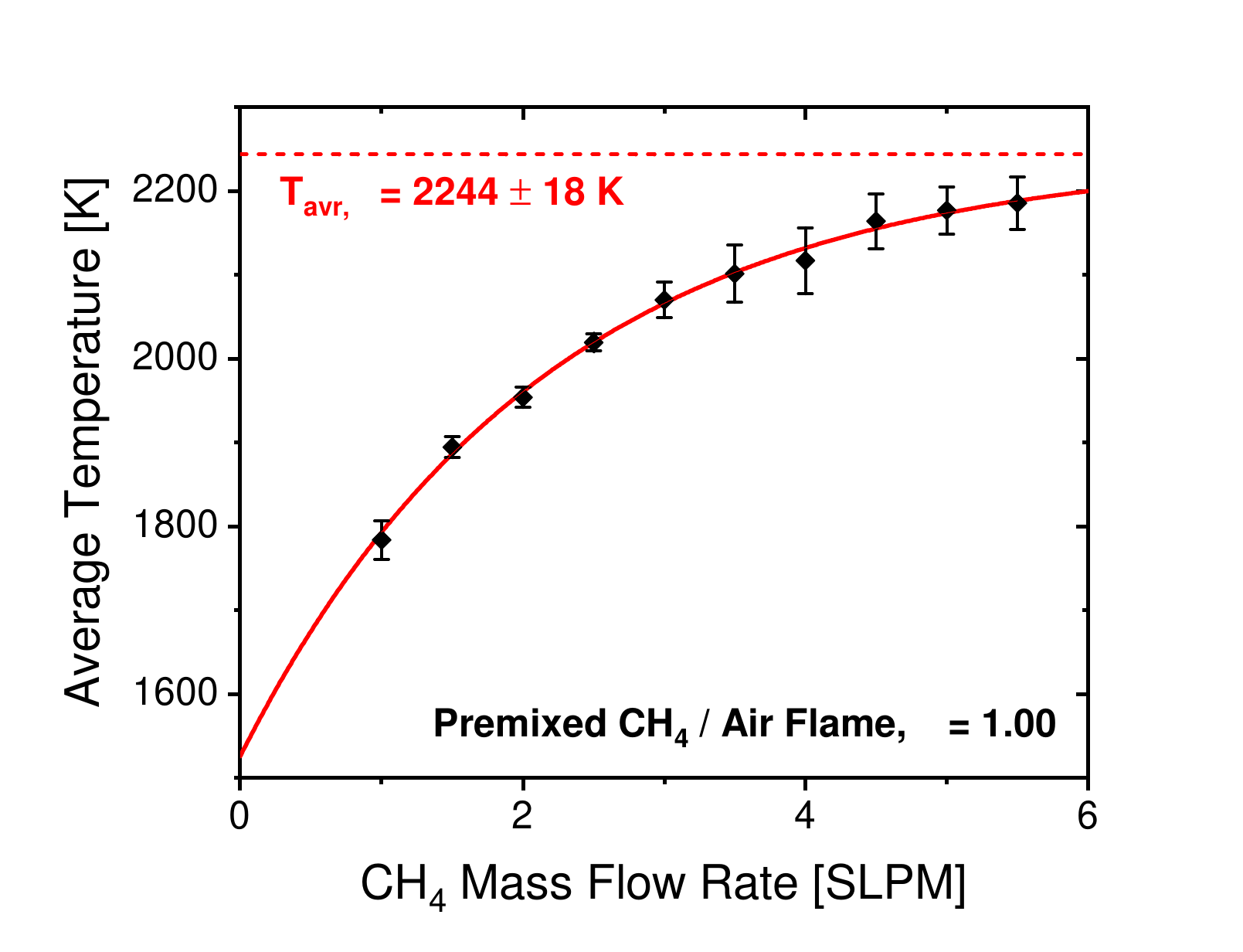}
\caption{Variation of $T_{core}$ as a function of $\dot{m}_{CH_4}$ at $\phi = 1.00$. The extrapolated value of $T_{core}$ at infinite $\dot{m}_{CH_4}$ agrees with the theoretical value of the adiabatic flame temperature. The error bars represent the standard deviation of the temperature distribution along the radial direction.}
\label{Fig6}
\end{figure}

\begin{figure}[ht!]
\centering
\includegraphics[width=0.5\linewidth]{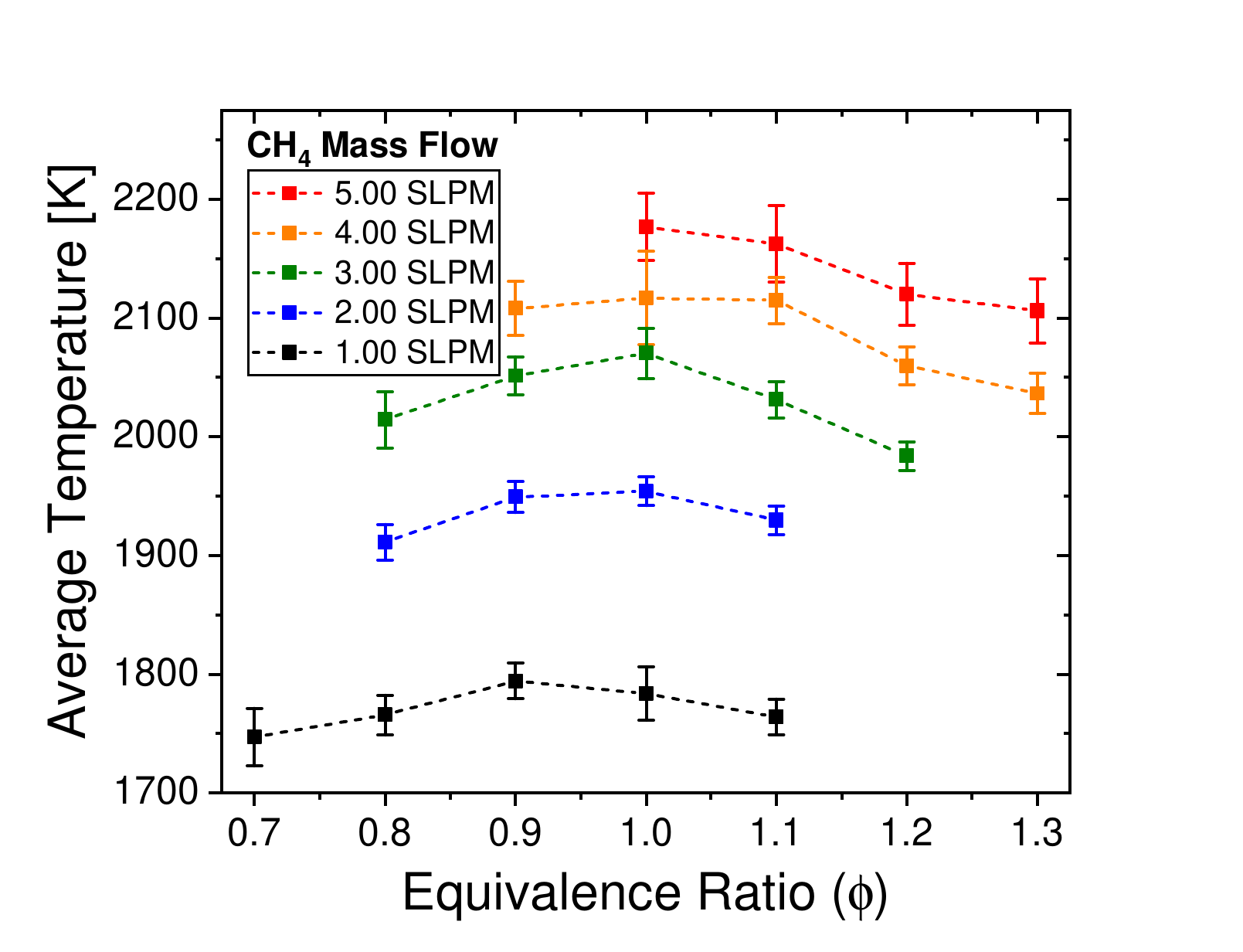}
\caption{$T_{core}$ measured at different $\phi$ and $\dot{m}_{CH_4}$.}
\label{Fig7}
\end{figure}

In the core region of the flame (defined as a central circle with a radius of 20 mm), the average mole fraction of $\rm CO_2$ was determined to be $X_{core} = 8.84 \pm 0.08 \%$, and the average temperature (weighted by the $\rm CO_2$ mole fraction) was $T_{core} = 2020 \pm 10$ K. This value was lower than the adiabatic flame temperature ($T_{ad}$ =  2229 K, as calculated based on the FFCM-2 mechanism \cite{ZDV2023}), due to heat loss near the burner surface and possibly the diffusion of the shielding nitrogen into the core flame. Experiments at different flow rates but the same equivalence ratio were also conducted, and the results (shown in Fig. \ref{Fig6}) indicated that $T_{core}$ approached $T_{ad}$ asymptotically with increasing $\rm CH_4$ mass flow rate ($\dot{m}_{CH_4}$). Based on an exponential extrapolation, the core temperature at infinite $\dot{m}_{\rm{CH_4}}$ was estimated to be $T_{\infty} = 2244 \pm 18$ K, which agreed with the theoretical value of $T_{ad}$ within its uncertainty range. Additional experiments were also performed under different conditions, and the results are summarized in Table \ref{Table1} and Fig. \ref{Fig7}.

\begin{table}[h!] \footnotesize
\caption{Core flow temperature and CO$_2$ concentration of axial-symmetric flat flames}
\centerline{\begin{tabular}{c c c c c c c}
\hline 
Case \# & $\phi$ & CH$_4$ flow rate & Air flow rate & N$_2$ flow rate & $T_{core}$ & $X_{core}$ \\
& & [SLPM] & [SLPM] & [SLPM] & [K] & [\%]\\
\hline
1 & 0.70 & 1.00 & 13.83 & 15.69 & 1747 $\pm$ 24 & 6.62 $\pm$ 0.20 \\
\hline
2 & 0.80 & 1.00 & 12.09 & 13.84 & 1766 $\pm$ 17 & 7.79 $\pm$ 0.13 \\
3 & 0.80 & 2.00 & 24.18 & 27.69 & 1911 $\pm$ 15 & 7.55 $\pm$ 0.17 \\
4 & 0.80 & 3.00 & 36.27 & 41.53 & 2014 $\pm$ 24 & 7.61 $\pm$ 0.13 \\
\hline
5 & 0.90 & 1.00 & 10.75 & 12.42 & 1794 $\pm$ 15 & 8.65 $\pm$ 0.16 \\
6 & 0.90 & 2.00 & 21.49 & 24.84 & 1949 $\pm$ 13 & 8.47 $\pm$ 0.13 \\
7 & 0.90 & 3.00 & 32.24 & 37.26 & 2051 $\pm$ 16 & 8.46 $\pm$ 0.14 \\
8 & 0.90 & 4.00 & 42.98 & 49.69 & 2108 $\pm$ 23 & 8.23 $\pm$ 0.16 \\
9 & 0.90 & 4.51 & 48.46 & 56.02 & 2149 $\pm$ 33 & 8.26 $\pm$ 0.18 \\
\hline
10 & 1.00 & 1.00 & 9.66  & 11.27 & 1784 $\pm$ 23 & 8.74 $\pm$ 0.25 \\
11 & 1.00 & 1.50 & 14.49 & 16.90 & 1894 $\pm$ 12 & 8.93 $\pm$ 0.10 \\
12 & 1.00 & 2.00 & 19.32 & 22.54 & 1954 $\pm$ 12 & 8.86 $\pm$ 0.16 \\
13 & 1.00 & 2.50 & 24.15 & 28.17 & 2020 $\pm$ 10 & 8.84 $\pm$ 0.08 \\
14 & 1.00 & 3.00 & 28.99 & 33.80 & 2070 $\pm$ 21 & 8.81 $\pm$ 0.16 \\
15 & 1.00 & 3.50 & 33.82 & 39.44 & 2101 $\pm$ 34 & 8.68 $\pm$ 0.31 \\
16 & 1.00 & 4.00 & 38.65 & 45.07 & 2117 $\pm$ 39 & 8.51 $\pm$ 0.30 \\
17 & 1.00 & 4.50 & 43.48 & 50.71 & 2164 $\pm$ 32 & 8.48 $\pm$ 0.22 \\
\hline
18 & 1.10 & 1.00 & 8.78  & 10.34 & 1764 $\pm$ 15 & 8.82 $\pm$ 0.22 \\
19 & 1.10 & 2.00 & 17.56 & 20.68 & 1930 $\pm$ 12 & 8.60 $\pm$ 0.13 \\
20 & 1.10 & 3.00 & 26.34 & 31.02 & 2031 $\pm$ 15 & 8.28 $\pm$ 0.10 \\
21 & 1.10 & 4.00 & 35.12 & 41.36 & 2115 $\pm$ 19 & 8.38 $\pm$ 0.16 \\
22 & 1.10 & 5.00 & 43.90 & 51.70 & 2162 $\pm$ 32 & 8.02 $\pm$ 0.19 \\
\hline 
23 & 1.20 & 3.00 & 24.14 & 28.70 & 1984 $\pm$ 12 & 7.24 $\pm$ 0.11 \\
24 & 1.20 & 4.00 & 32.19 & 38.26 & 2060 $\pm$ 16 & 6.94 $\pm$ 0.07 \\
25 & 1.20 & 5.00 & 40.24 & 47.83 & 2120 $\pm$ 26 & 7.15 $\pm$ 0.14 \\
26 & 1.20 & 6.00 & 48.28 & 57.39 & 2178 $\pm$ 37 & 6.91 $\pm$ 0.22 \\
\hline 
27 & 1.30 & 4.00 & 29.74 & 35.68 & 2037 $\pm$ 17 & 5.86 $\pm$ 0.07 \\
28 & 1.30 & 5.00 & 37.18 & 44.60 & 2106 $\pm$ 27 & 5.74 $\pm$ 0.14 \\
\hline 
\end{tabular}}
\label{Table1}
\end{table}

\begin{figure}[h!]
\centering
\includegraphics[width=0.5\linewidth]{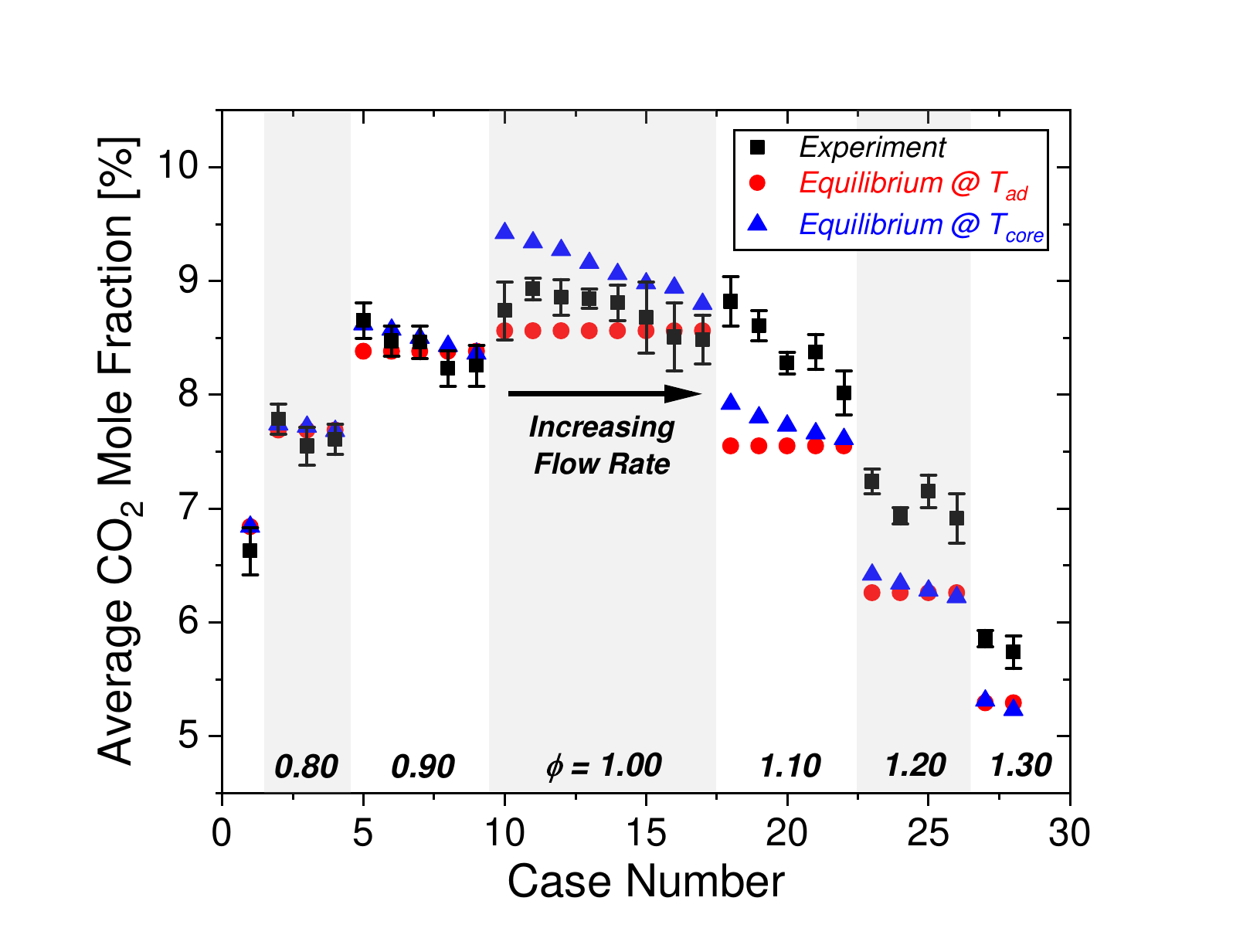}
\caption{Comparison of the measured CO$_2$ mole fractions with the equilibrium values for all cases in Table 1. Error bars represent the 1$\sigma$ measurement uncertainties.}
\label{Fig8}
\end{figure}

For all cases listed in Table 1, the measured CO$_2$ concentrations were compared with their local equilibrium values, as illustrated in Fig. \ref{Fig8}. At fuel-lean conditions, the measured CO$_2$ concentrations agreed well with the equilibrium values. At stoichiometric conditions, the measured values were slightly higher than the equilibrium values at adiabatic flame temperatures and slightly lower than the equilibrium values at the measured temperatures. At fuel-rich conditions, the measurements were consistently higher than the equilibrium values at both temperatures. These deviations were likely caused by entrainment of ambient air into the flames, which reduced the actual equivalence ratios and increased the CO$_2$ concentrations beyond the nominal equilibrium values. The presence of ambient air entrainment is also supported by the temperature measurements at rich conditions, where the measured flame temperatures were also slightly higher than the adiabatic flame temperatures at the nominal equivalence ratios. Such entrainment effects can also explain the differences between the measured and equilibrium CO$_2$ concentrations at stoichiometric conditions. Nonetheless, these differences diminish at high flow rates, where the fraction of entrained air asymptotically approaches zero.

\subsection{Volumetric Thermometry of Axisymmetric Flat Flames}
\begin{figure}[h!]
\centering
\includegraphics[width=\linewidth]{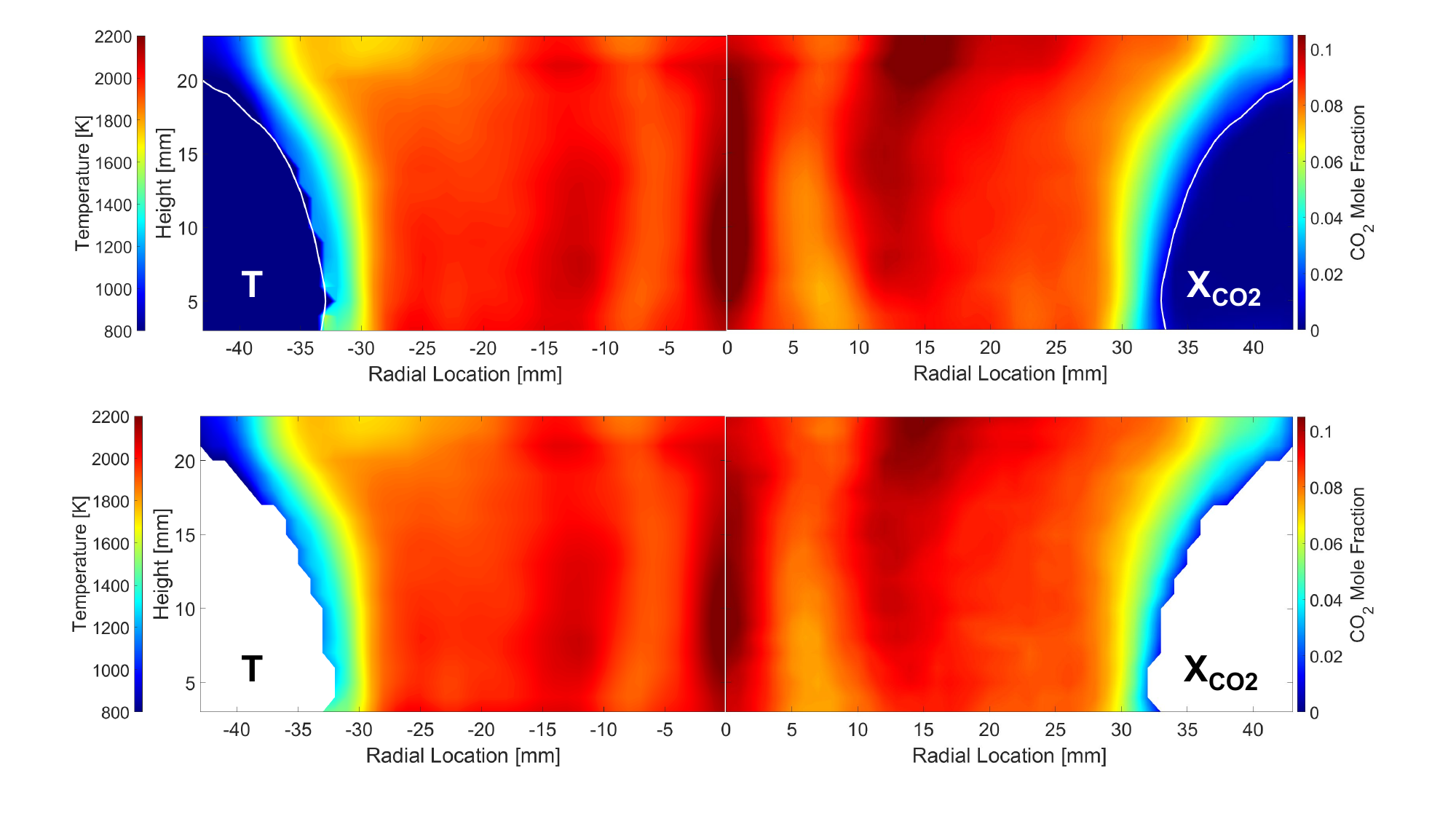}
\caption{$\boldsymbol{T}$ (left) and $\boldsymbol{X}$ (right) in a single-shot volumetric measurement of an axisymmetric flame. Conditions are similar to those in Fig. \ref{Fig4}. Top panel: results of the 1st-round iteration, with the white lines indicating the contour of $X_{CO_2} = 0.01$. Bottom panel: final results of the full iterations.}
\label{Fig9}
\end{figure}

\begin{figure}[h!]
\centering
\includegraphics[width=\linewidth]{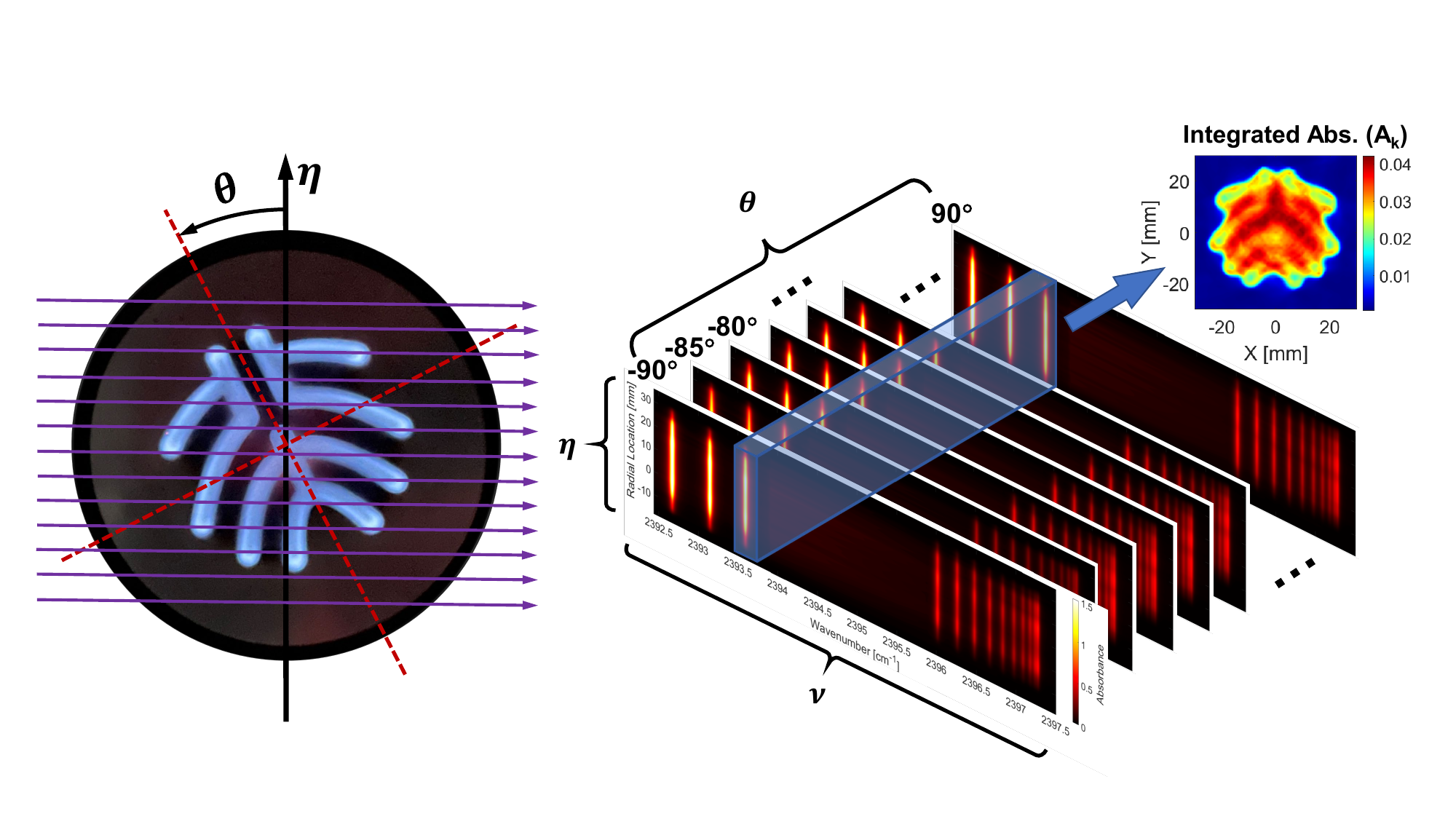}
\caption{A schematic view of the PKU flame measurement and its data structure.}
\label{Fig10}
\end{figure}

Fig. \ref{Fig9} shows an example set of results obtained from a single-shot volumetric measurement of an axisymmetric flame under conditions similar to those in Fig. \ref{Eqn4}. The laser beam was scanned over a rectangular domain of 50 mm $\times$ 25 mm in the vertical plane, spanning from -5 mm to 45 mm in the radial direction and from 2.5 mm to 22.5 mm in the height direction. Shown in the top panel of the figure are $\boldsymbol{T}$ and $\boldsymbol{X}$ inferred from a first round of iterations using Eqn. \ref{Eqn10}. A spatial mask was then applied to the points of $X_{CO_2} < 0.01$, which accelerated the remaining iterations. The final results of $\boldsymbol{T}$ and $\boldsymbol{X}$ are shown in the bottom panel of Fig. \ref{Fig10}. The flame was stretched due to stagnation at a height of 30 mm above the burner, and strong diffusion/mixing with the shielding nitrogen was observed in a zone approximately 3-5 mm thick near the edge of the flame. In spite of some local hot/cool plumes, the temperature distribution in the core region was relatively uniform.

\subsection{Planar Thermometry of the PKU Flame}
The beam pattern and data structure of the PKU flame measurement are illustrated in Fig. \ref{Fig10}. Throughout the course of this measurement, the flame remained stable and was rotated from $-90^\circ$ to $90^\circ$ in increments of $5^\circ$, whereas the laser beams were at a fixed angle. At each angle, the beam position was scanned from -25 mm to 25 mm in the radial direction, and at each beam position, the laser wavenumber was swept between 2392.45 $\rm{cm^{-1}}$ and 2397.50 $\rm{cm^{-1}}$. The resulting data were stored as a 3-D matrix.

Similar to the previous example, an initial estimate of $\boldsymbol{T}$ and $\boldsymbol{X}$ was performed to accelerate the data analysis. The spectrally integrated absorbance at each point in the measurement plane, $A_k(\boldsymbol{x_i})$, was first calculated using the inverse Radon transform based on the line-of-sight data at different beam positions and view angles. The gas temperature and $\rm CO_2$ concentration were then estimated point by point from the integrated absorbance of different transition features. After that, a spatial mask was applied to points where $X_{CO_2} < 0.02$. These estimated results, as shown in the top panels of Fig. \ref{Eqn10}, were used as the initial values in the iterative solution of Eqns. \ref{Eqn9} and \ref{Eqn10}.

\begin{figure}[h!]
\centering
\includegraphics[width=\linewidth]{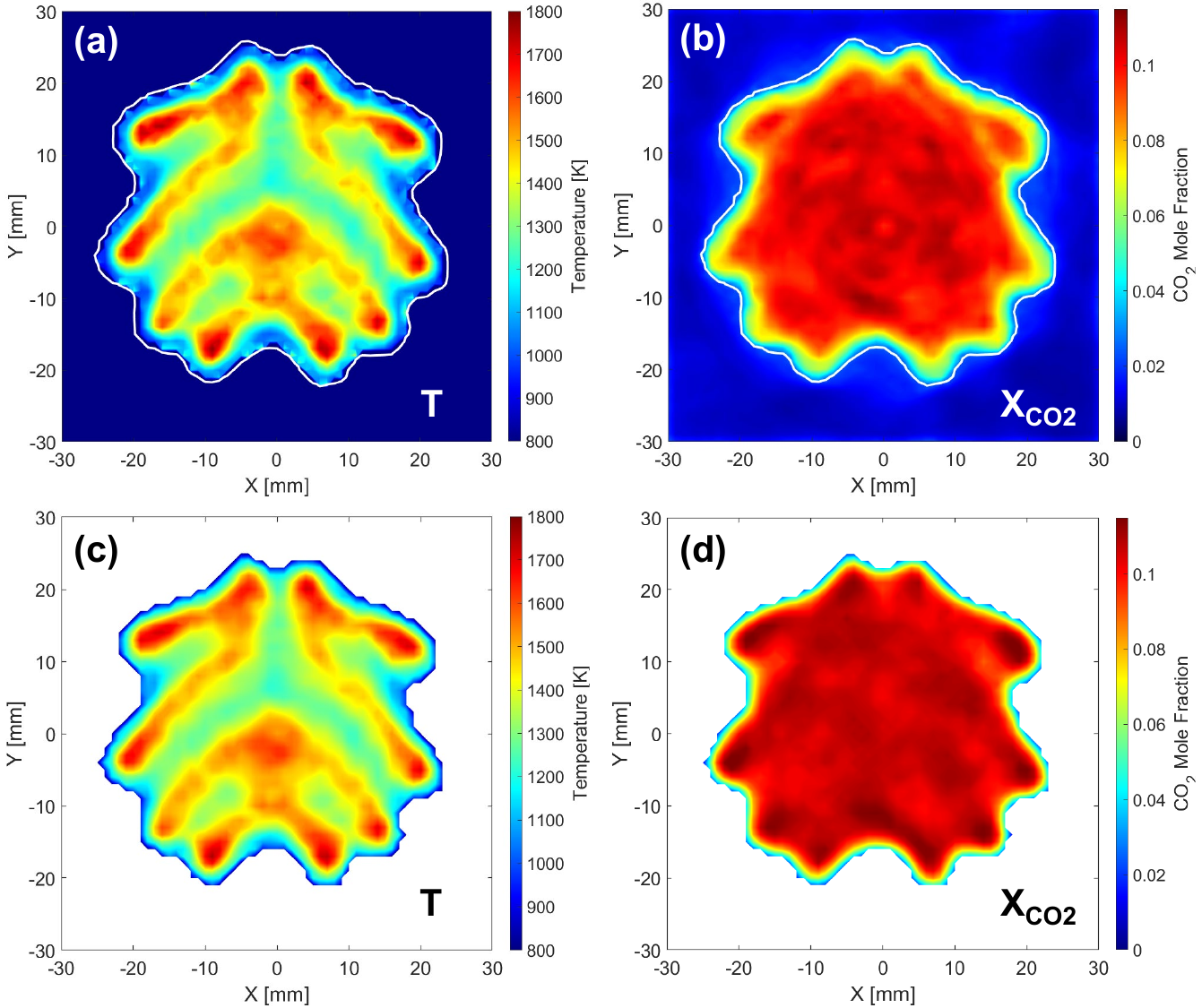}
\caption{Results of the $\boldsymbol{T}$ (left) and $\boldsymbol{X}$ (right) measurements in the PKU flame at $\dot{m}_{C_3H_8}$ = 0.40 SLPM, $\dot{m}_{air}$ = 9.61 SLPM, and $\phi$ = 1.00. Top panels: initial estimate based on the inverse Radon transform, with the white lines indicating the contour of $X_{CO_2} = 0.02$. Bottom panels: final results of the iterative inference.}
\label{Fig11}
\end{figure}

The final results of $\boldsymbol{T}$ and $\boldsymbol{X}$ in the PKU flame are shown in the bottom panels of Fig. \ref{Fig11}. An excellent level of resolution, precision, and contrast was attained by the current thermometry method. The inferred gas temperature was seen to agree well with the expected distribution based on the chemiluminescence image shown in Fig. \ref{Fig10}, with hot zones aligned with the contour of the PKU logo (where the flame fronts were located) and extended warm regions encompassing the outer boundary of the flame. The $\rm CO_2$ mole fraction distribution was found to be relatively uniform over the entire region of interest, likely due to the strong convection and expansion of combustion products from the flame fronts to the warm regions.

\section{Conclusions}
In the current study, an accurate and robust method for spatially resolved measurements of gas temperatures in flames and reacting flows was developed using multi-color $\rm CO_2$ laser absorption spectroscopy. This method utilized a collinear dual-laser configuration to access tens of CO$_2$ absorption transitions between 2392.45 $\rm{cm^{-1}}$ and 2397.50 $\rm{cm^{-1}}$ every 100 $\rm{\mu s}$. From their relative intensities, the gas temperatures (as well as the CO$_2$ concentrations) were determined with high sensitivity and robustness. A high-speed 2-D parallel beam scanning system was introduced to achieve a planar field measurement speed of 200 Hz and a volumetric field measurement speed of 2 Hz, at an effective spatial resolution of 1 mm. Additionally, a physically constrained iterative inference framework was developed for the quantitative analysis of the measurement data.

A series of proof-of-concept laminar flame experiments demonstrated a high level of accuracy (1\% for typical single-shot measurements), precision, resolution, and contrast of the current thermometry method. For measurements at lower temperatures and in gases with relatively simple compositions, the performance can also be validated using alternative approaches, such as heated gas cells or shock waves. Further improvements in measurement speed (by an order of magnitude), as well as extensions to high-pressure measurements (via incorporating more transitions) and validations in other forms of flame experiments, are currently in progress. Given its high performance metrics and relative ease of use, this thermometry method, along with its data analysis framework, promises to provide significant utility in future combustion studies.

\section*{Acknowledgments}
This research is supported by the National Natural Science Foundation of China under Grants No. 92152108 and No. 12472278. 

\bibliographystyle{elsarticle-num-names}
\bibliography{References}

\end{document}